# Nanomaterials for Supercapacitors: Uncovering Research Themes with Unsupervised Machine Learning

MridhulaVenkatanarayanan[1], Amit K Chakraborty[2], Sayantari Ghosh[2,*]

## Abstract

Identification of important topics in a text can facilitate knowledge curation, discover thematic trends, and predict future directions. In this paper, we aim to quantitatively detect the most common research themes in the emerging supercapacitor research area, and summarize their trends and characteristics through the proposed unsupervised, machine learning approach. We have retrieved the complete reference entries of article abstracts from Scopus database for all original research articles from 2004 to 2021. Abstracts were processed through a natural language processing pipeline and analyzed by a latent Dirichlet allocation topic modeling algorithm for unsupervised topic discovery. Nine major topics were further examined through topic-word associations, Inter-topic distance map and topic-specific word cloud. We observed the greatest importance is being given to performance metrics (28.2%), flexible electronics (8%), and graphene-based nanocomposites (10.9%). The analysis also points out crucial future research directions towards bio-derived carbon nanomaterials (such as RGO) and flexible supercapacitors.

.

## 1. Introduction:

Our constant endeavours to enhance the quality of life have paved the way for the staggering The demand for energy has seen a steep rise in the recent years due to our increasing reliance on technologies to improve our quality of living. In addition to developing next-generation sustainable energy technologies, the industries and consumers pose an increasing demand for low mass electrochemical devices offering high energy and power density. Supercapacitors are electrochemical energy storage devices that are considered to have enormous potential to bridge the gap between rechargeable batteries and dielectric capacitors since these can exhibit high power density (1-10 kW $kg^{-1}$) and long cycle life (~$10^5$ cycles) [1, 2]. Owing to an exponential demand for energy storage systems offering high power output in the industrial, residential, and transportation sectors, the sales revenue of the supercapacitor industry has spiked from $40 million in 1989 to $877 million in 2014 [3]. Consequently, supercapacitor research and development is becoming one of the most discussed topics in the research fraternity [4, 5, 6, 7].

[1]Department of Metallurgical and Materials Engineering, National Institute of Technology Trichy
[2] Department of Physics, National Institute of Technology Durgapur

* Corresponding author

In contrast to the previously used trial and error-based approach for material discovery, the rapid advancement of high-performance computing has expedited the discovery of structure-property relationships. The advent of state-of-the-art supercomputers and sophisticated algorithms has enabled quantum-mechanical approaches such as Density Functional Theory to populate material databases such as Materials Project (MP) database[8] and Open Quantum Mechanical Database (OQMD)[9]. While this is advantageous, the unrealistically high computational cost has bottlenecked the transferability of such Quantum Mechanical methods to complex systems[10]. The emergence of Artificial Intelligence has brought about a paradigm shift in material discovery by unearthing the intricate structure-property relationships from material databases [11, 12].

The prediction of trends and properties using Machine Learning (ML) dictates the need to have highly structured databases. Needless to say, with more than 27,782 research articles published in the span of the last 20 years that explored the material research related to supercapacitors [13], these publications have a cornucopia of a large amount of unstructured knowledge present in the form of text and images. When carefully collected and curated in the form of a database, it can be subjected to ML techniques to offer valuable insights regarding the underlying structure-property relationships. Manual creation of such databases from images and text, in addition to being laborious and time-consuming, also requires sound domain expertise to comprehend and agglomerate the knowledge in an organized fashion. The contemporary research in the field of Natural Language Processing offers a roundabout to the problems imposed by manual inspection. NLP attempts to automate the interpretation of textual data by transforming the written text into a form that computers can easily manipulate. [14]

While NLP has been used widely for solving the challenges pertaining to biological sciences, it has only recently found its application in addressing the problems faced by the material science community, making it a hot research target for contemporary researchers. Shu et al.[15] have created an extensive database of battery materials and their associated properties using the toolkit ChemDataExtractor which aided in the automated extraction of data from existing literature. They have also used ChemDataExtractor integrated with an improvised Snowball algorithm to extract Curie and Neel temperatures of various chemical compounds. Venugopal et al.[17] have demonstrated the usage of Natural Language Processing techniques such as topic modeling and caption cluster plots to answer specific questions related to the synthesis and characterization of inorganic glasses. Shetty et al.[18] have analyzed the trends in the application of polymers over the years using word vectors.

In NLP, topic Modeling is an unsupervised machine learning technique that automatically discovers the dormant underlying topics in a corpus based on the principle that a corpus is a collection of documents comprising several topics. Each topic in itself is a probabilistic distribution of words. Latent Dirichlet Allocation, a generative probabilistic model of a corpus, is a topic modeling technique introduced by Blei et al.[19] based on the idea that documents comprise several latent topics, and each topic can be represented by a random mixture of words such that the words belong to the topic with some probability distribution. Topic modeling methods such as LDA can offer a treasure trove of knowledge when applied to material science literature, which could help solve open challenges such as novel material discovery. [17] Though these tool can help uncover niche topics with the potential for further exploration, contributing to a shift in research dynamics, this is a field mostly unexplored. In recent times, Huo et al.[20] combined LDA and Random Forest to cluster words related to specific experimental material synthesis steps in one such work.

In our study, we focus on the nanomaterials that are being predominantly used as supercapacitor electrode materials with an unsupervised learning approach. We attempt to perform topic modeling on the corpus of abstracts corresponding to the literature on supercapacitors and nanomaterials. Our primary goal is to cluster together the characterization and functionality information corresponding to electrode materials using LDA. Further, we attempt to identify the most researched topics corresponding to these materials. In addition to revealing essential aspects of the state-of-the-art supercapacitor research, this work also strives to offer a bird's eye view of the research practices in this domain, which will be beneficial in contributing to the novel supercapacitor material discovery. The major contributions of this work are:

- To compile an NLP-compatible database using data mining and pre-processing for consolidating major aspects captured in the developing and rapidly growing literature on Supercapacitors
- To provide an in-depth thorough organization of the research conditions and trends imposed on materials related to supercapacitors,
- To offer an overall visualization of the research practices in this domain which could help researchers to obtain personalized assessment of the state-of-the-art practices.

The rest of this article is organized as follows: In Section 2, we will discuss the methodology in detail, keeping the focus on Text mining, text retrieval, pre-processing, performing LDA on the corpus, and visualization of the results. In section 3, results will be discussed, with an elaborate discussion on major topics. In section 4, we point out future directions and conclude.

## 2. Methodology

The proposed pipeline contains four sequential sub-modules to identify the distinct clusters from the raw text. A schematic diagram of the method pipeline used is shown in Fig. 1, and the steps are elaborately discussed in the upcoming subsections.

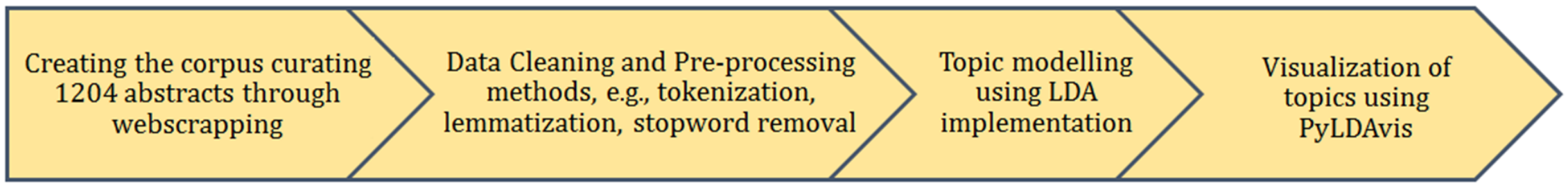

Fig. 1: A schematic diagram of the methodology pipeline

### A. Text Retrieval

A corpus of the abstracts corresponding to 1204 research articles was created using the Elsevier Science Direct API. The choice of scraping only the abstracts and not the entire research article for the creation of corpus is justified as abstracts provide a holistic view of the whole research paper by condensing the critical aspects of the study such as the research questions, material under investigation, synthesis routes, experimental methodology, characterization techniques, metrics, trends, and results. The corpus corresponding to the inputted queries "Nanomaterials" and "Supercapacitors" was scrapped and collected in a CSV file[2]. Abstracts of research articles containing the terms "Supercapacitors" and "Nanomaterials" published during the timeframe ranging between 2004 and 2021 were curated.

B. Data cleaning and pre-processing:

The abstracts stored in the CSV file are subjected to pre-processing techniques before topic modeling is done. Initially, the redundant abstracts are filtered out, and the corpus consists of 1182 unique abstracts. A series of 'Regular Expression' functions are defined to remove the punctuations, accented and special characters, alphanumerical characters, digits, additional spaces, units associated with the measurements, and roman numerals. Following this, the words are converted into lowercase, and further operations such as tokenization, removal of stopwords, and lemmatization are performed. Tokenization is a process that takes up the white space within the unstructured raw text as a delimiter and breaks it up into individual words. Stopwords are the frequently occurring words in the language and are removed to offer a spotlight to the essential words in the corpus that add meaning to the textual data. Stemming and lemmatization are operations that enhance the model's performance by condensing a word into its root form. Unlike stemming, which leads to data loss, lemmatization takes the word's context into account and returns the dictionary form of the word. For the reason mentioned above, this study has preferred lemmatization over stemming. The Natural Language Processing Toolkit (NLTK) has been imported into python to perform tokenization, stopwords removal, and lemmatization operations. Following this, the words co-occurring frequently are made into a single term by importing libraries such as bigrams and trigrams provided by Gensim. A few examples of the words frequently co-occurring in the corpus include "specific_capacitance," power_density," "electrochemical_impedance_spectroscopy," "transmission_electron_microscopy," etc.

C. Latent Dirichlet Allocation

A dictionary is created by assigning each unique word in the collection of documents to an index. This dictionary of unique words is subsequently used to create the document-term matrix of the corpus (Bag of words representation). The bag of words representation assigns each word with a vectorial representation of fixed length based on the word's frequency. A bag of words based approach is chosen to represent the corpus to impart the model with the flexibility of overlooking the order of words and documents. This implies that the model does not take the order of the words in the document as well as the order of documents in the corpus into consideration. [(21)] The parameters such as document-term matrix and dictionary are subsequently used by Gensim's implementation of LDA to train the topic model. In this method, the posterior probabilities fora document collection determine a decomposition of the collection into topics. Suppose in a corpus $C$ with $N$ topics; we have $A$ abstracts with up to $K$ number of word tokens each from a vocabulary $M$. In that case, each response has an $N$-dimensional multinominal distribution $\phi_d$ over topics with a common Dirichlet prior Dir($\delta$). Each topic has an M dimensional multinominal $\delta_N$ over words with a common symmetric Dirichlet prior Dir($\mu$). To estimate $\delta$, $\mu$ from a corpus **C**, we maximize the log-likelihood ln P(C|$\delta$, $\mu$). We have used 'gensim' and 'nltk' packages to implement LDA on our responses.

[2] Link provided for data availability:
https://drive.google.com/file/d/13B_l_t1wllDTbIGKN0I8pbM94YYauf5X/view

We compared all of the models side by side from 3 topics to 20 topics using their default parameters with our dataset. We chose 10 to be the optimum number of topics because we wanted as many topics as possible without reducing model quality and without the topics becoming too narrow. Further fine-tuning of the model was performed to choose the best initial parameters for consensus judgement of the authors after several iterations of proposals. While there have been attempts at automatic topic naming, there are reports of those being imprecise, having no guarantee of coherence[(22)].To produce meaningful topic names, we decided to manually name each topic based on experience with experimental research of the authors and the content of the most important 20 words within each topic.

All the analysis was performed using PYTHON 3. The Natural Language Processing Toolkit (NLTK) has been imported into python to perform pre-processing operations such as tokenization, stopwords removal, and lemmatization. The Gensim Python Library's implementation of LDA is used in this study to train the model. Gensim is used to extract the semantic structure of documents automatically[(23)]. Gensim's class Phrases[(24)] is used to detect bigrams and trigrams.

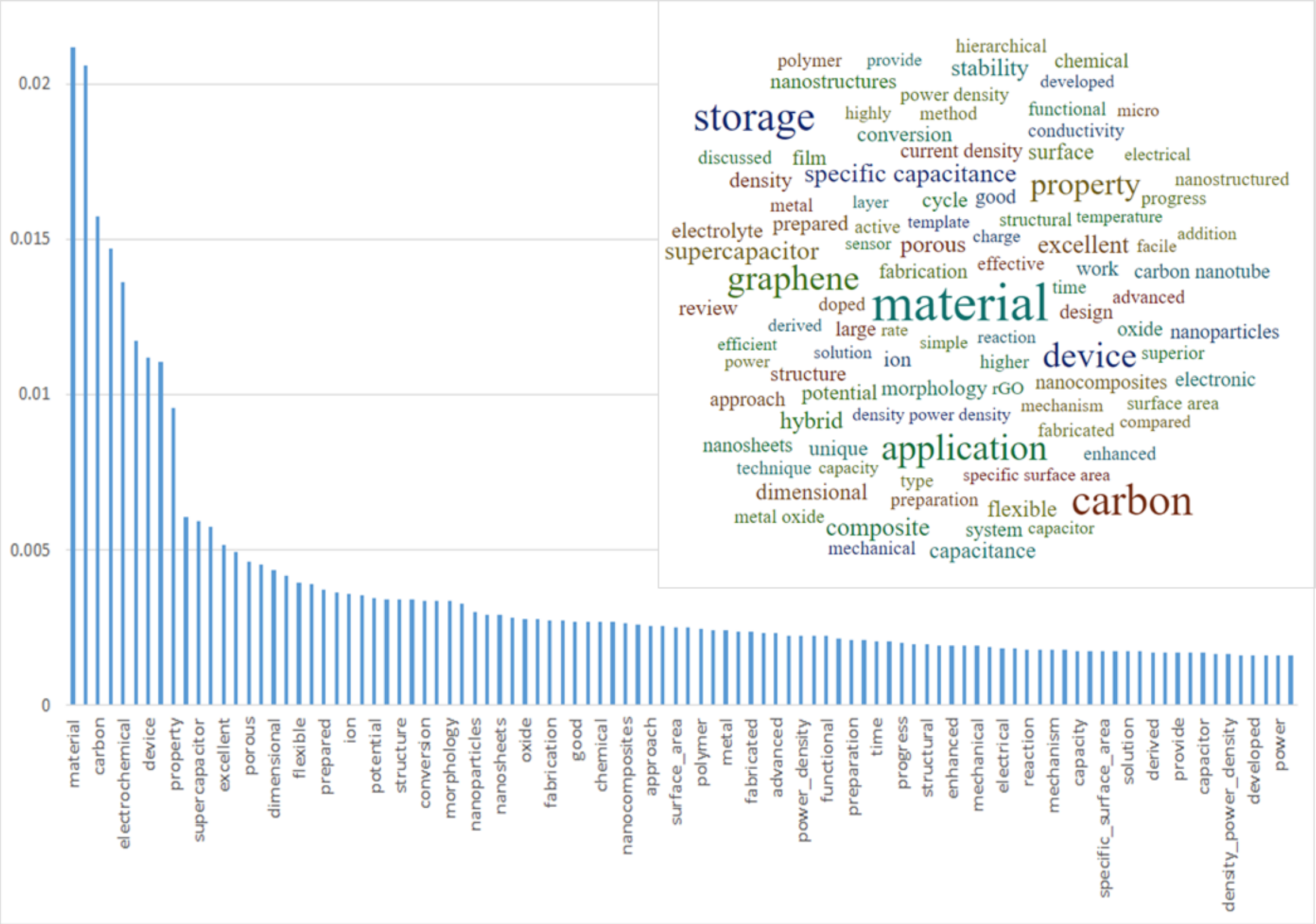

Figure 2: A preliminary report of TF analysis. The word-cloud is shown for visualization of most frequent 100 words of the corpus, while the size is dictated by the frequency. Absolute values of the frequencies can be estimated from the histogram.

### D. Visualization of the model

The visualization of the topic modeling was done with PyLDAvis. This interactive Python library is employed to visualize the topics generated using the topic model. In this visualization, we could separate the topics into distinct clusters when embedded in two-dimensional space and observe the exhibited consistent patterns of interaction. PyLDAvis provides two visualization panels. The circle area shows the significance of the particular topic in the entire corpus. The distance between the circles is a measure of the similarity between the topics. The right side of the panel lists the 30 most frequently occurring words corresponding to each topic. The blue bars depict the total frequency of a particular word in the entire corpus. The red bar represents the frequency of a given word in a specific topic.

## 3. Results

A. Preliminary Analysis using Term Frequency Calculation

Term frequency (TF) analysis, usually visualized by word-clouds and histograms, is a basic method to estimate the importance of a given word in the corpus by the frequency of occurrence of that word. We calculate the TF, normalised by the total number of terms in the document, i.e., $TF(t) = \frac{n_t}{N_d}$, where, $n_t$ is the frequency of term *t* in the document $d$ with $N_d$ number of total terms. In our corpus, 6138 different terms are there with the total number of terms being 81826. The result is shown in Fig. 2.

We note words like “material”, “electrode”, “graphene”, and “carbon” occur most frequently in the corpus for obvious reasons as the abstracts in the corpus talk extensively about the usage of carbon-based nanomaterials for storage applications, especially supercapacitor electrodes. Next set of words, like, “property”, “specific_capacitance”, “excellent”, “stability”, “conversion” etc. vaguely indicate certain performance parameters, without any specifically categorised output. This analysis points out that significance of a word might not be increasing proportionally with frequency. Moreover, a structured clustering might be able to categorize the different aspects on the research area, which is not possible by TF analysis.

B. Topic Modeling using Latent Dirichlet Allocation

As it is evident that TF analysis is not providing any clear and conclusive picture about the research directions, we proceed with our proposed methodology. Starting from 1204 identified articles, ten major topics covering characteristics and method-driven areas were discovered. The topics detected by unsupervised clustering through Latent Dirichlet Allocation, the method discussed in Section 2, are concisely reported in Table 1.

Table 1: Latent Dirichlet Allocation Topic Modeling Results.

| Topic number | Topic name | Keywords |
|---|---|---|
| 1 | Performance Metrics associated with supercapacitor electrode materials | Specific Capacitance, Current Density, Power Density, Capacitance Retention Cycle, Charge-Discharge, and Rate Capability |
| 2 | Importance of electrode materials in storage properties and applications | Material, Application, Storage, Property |
| 3 | Carbon based electrodes from biomass precursors. | Carbon, Biomass, Heteroatom, Dopant, Hollow, Template, Biomass-derived, Graphitization, Molten Salt, Quaternary, Carbonaceous, Nitrogen, Precursor, and Carbonization |
| 4 | Flexible, wearable, printable supercapacitors | Flexible, Miniatured, Volumetric Capacitance, Wearable electronics, Textile, Stretchable, Planar, Printable, Woven fiber, Flexible-wearable, Integrated, Interdigital |
| 5 | Graphene as electrode material | Graphene, Bioinspired, Methane, Nanoribbon(s) |
| 6 | Synthesis of Noble-metal stacked composite electrode | Ice, Stacked, Nucleation, Noble Metals, Charge-carrier |
| 7 | Morphology dependent Catalysis applications of various supercapacitor electrodes | Enriched, Microtubes, Conjoined, Heterostructures, Horseradish, Peroxidase, rGO, Porphyrin, Cocatalytic, Catalysed, Anodization, Banana, Anchovy, Oxygen Vacancy |
| 8 | Shapes of the electrode material | Nanoplatelets, Nanoneedle, Rod, Nanopowder, Microcubes |
| 9 | Reduced Graphene Oxide (rGO) as electrode material | rGO, Film |
| 10 | Author affiliations | Topic discarded |

The catalogued list of the thirty most frequently occurring words is visually inspected, and subsequently, a topic name is assigned by subsequent discussions among the authors. Because of several reported errors of automated methods for naming a topic, we avoided that and preferred the manual naming, despite being laborious. The Inter-topic Distance Map via a Multidimensional scaling plot generated by PyLDAvis is shown in Fig. 3, which provides an exhaustive view of the topic model. Each topic is represented by a circle, and the plot strives to give an outlook on the relevance of each of the individual topics with respect to the corpus and the similarity between each of the individual topics with respect to statistical nearness. The significance of each of the topics in the entirety of the corpus is represented by the area of the circle.

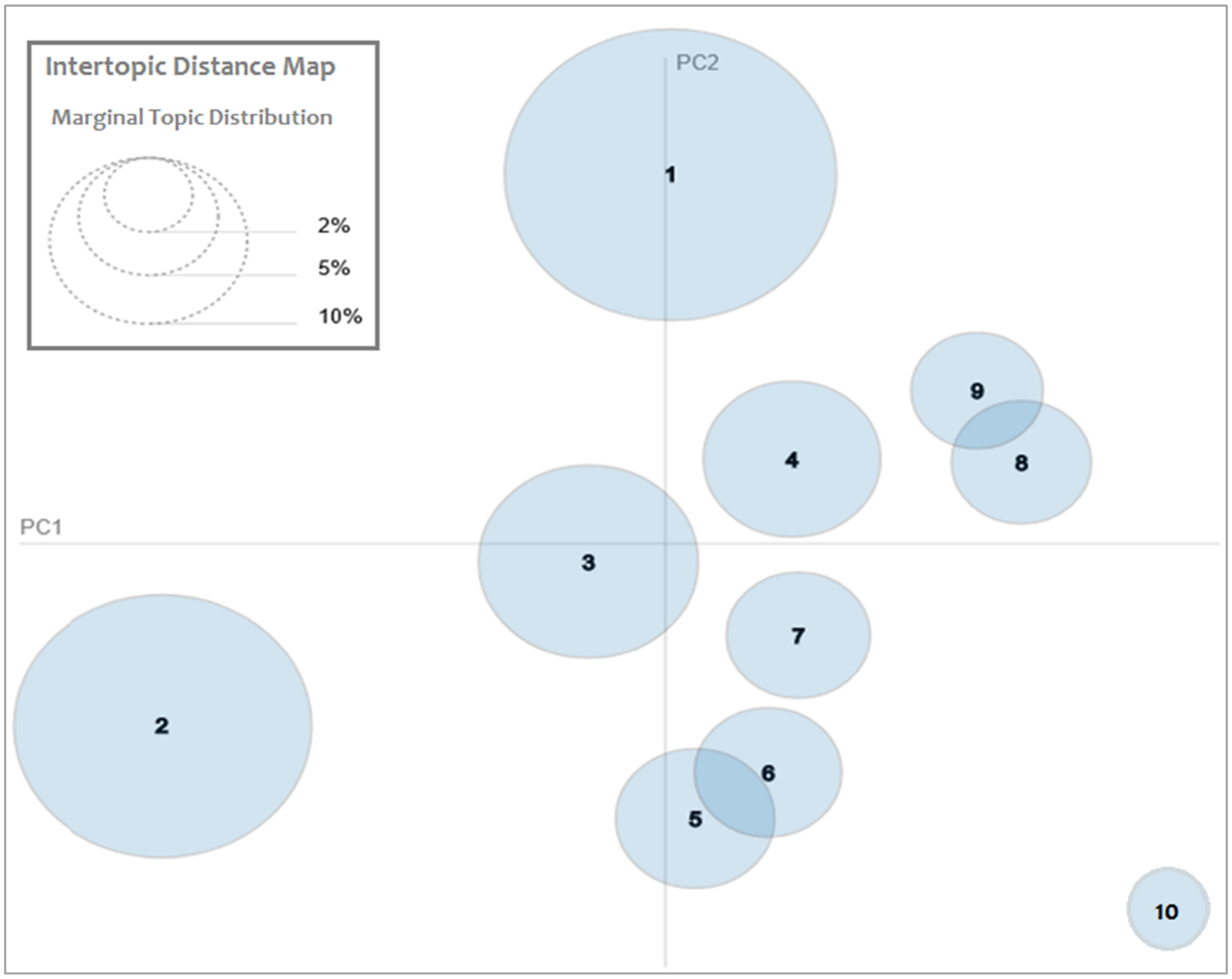

Figure 3: Inter-topic Distance Map via Multidimensional scaling plot for topics detected by Latent Dirichlet Allocation. Represented by circles in two dimensions, closely spaced topics have more similar words.

For an in-depth analysis of each topic, we first use the visualization readily available with PyLDAvis. It is a horizontal bar chart with each bar depicting a unique term corresponding to the topic. The relevance of a word to a particular topic is described by λ, *the lift*. The value associated with $0 \leq \lambda \leq 1$is a quantitative measure of the ratioof a word's frequency within a topic to its marginal probability across the corpus. So, with λ closer to 1,words are shown based on the occurrence throughout the corpus. On the other hand, the λ values closer to 0 display words based on the exclusivity of the word to the topic [(25)]. In this study, the optimal value of λ that assisted the best interpretation of topics from the listed terms is found to be 0.2. Associated bar charts and word clouds are shown in the sub-sections below, where we discuss in detail the major topics that uncover the current research themes and trends in the domain of supercapacitor research.

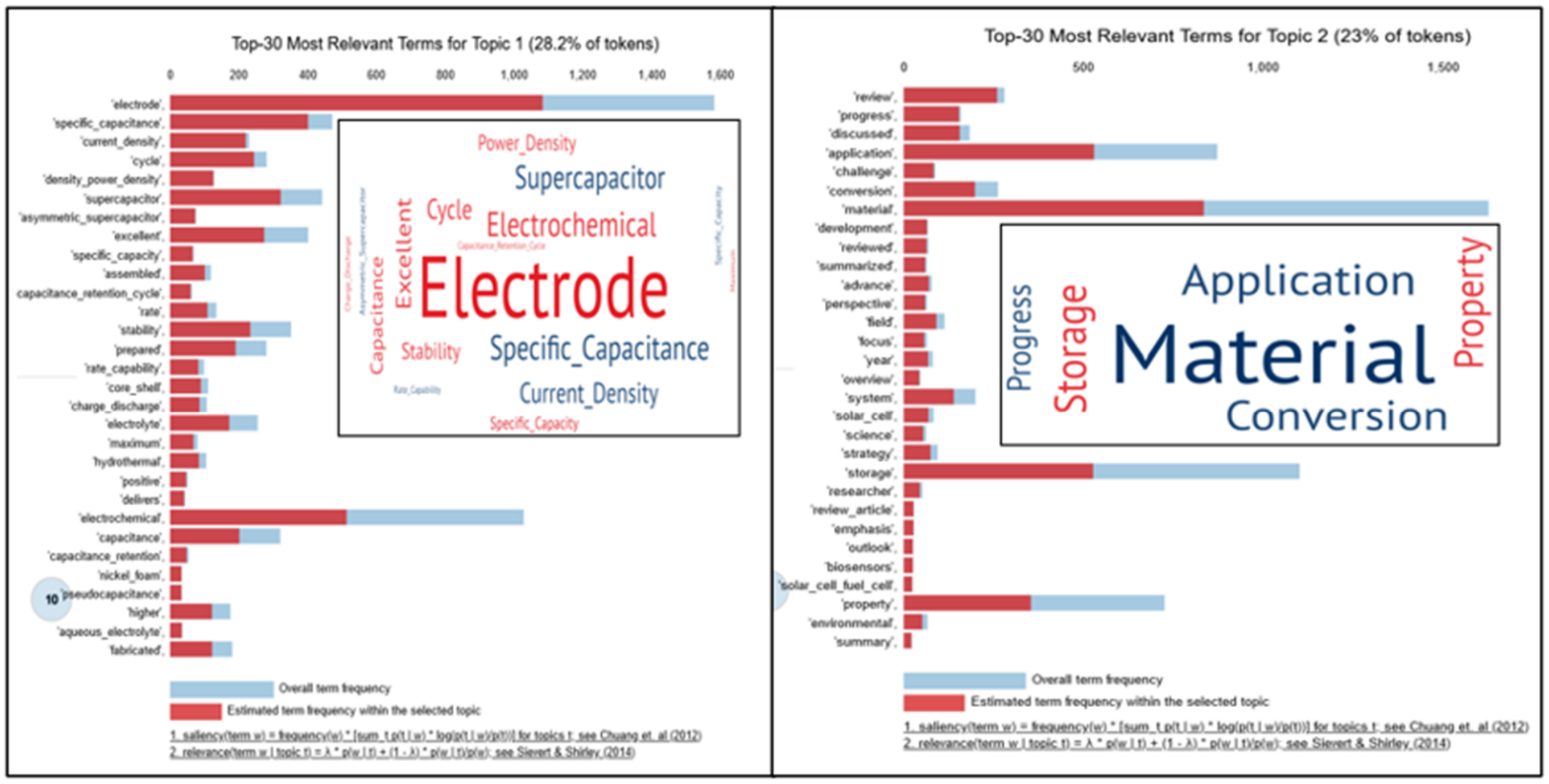

(a) (b)

Figure 4: Top words detected in (a) topic #3, (b) topic #4

## (1) Performance Metrics Associated with Supercapacitor Materials

As shown in Fig. 4(a), the majority of the words such as "specific capacitance," "current density," "power density," "capacitance retention cycle," "charge-discharge," and "rate capability" indicate that topic #1 corresponds to the performance metrics associated with supercapacitor materials. Terms such as "excellent," "maximum," and "higher" are used to quantify the degree of such parameters (Example: high power density, excellent rate capability, etc.). The term electrochemical hints at the term 'electrochemical performance' of electrode materials. Supercapacitors are next-generation energy storage devices offering an array of advantages such as high power density, high energy density, long cycle life, fast charge and discharge, instantaneous high current discharge, etc[26]. However, supercapacitors being bottlenecked by their low energy density are under scrutiny by the research fraternity, which justifies the presence of such performance parameters in our corpus.

### (2) Importance of electrode materials in storage properties and applications

The primary keywords of this topic being "Material," "Application," "Storage," and "Property" can be seen in Fig. 4(b). This suggests that this particular topic gives a brief overview of the importance of electrode materials in storage properties and applications.

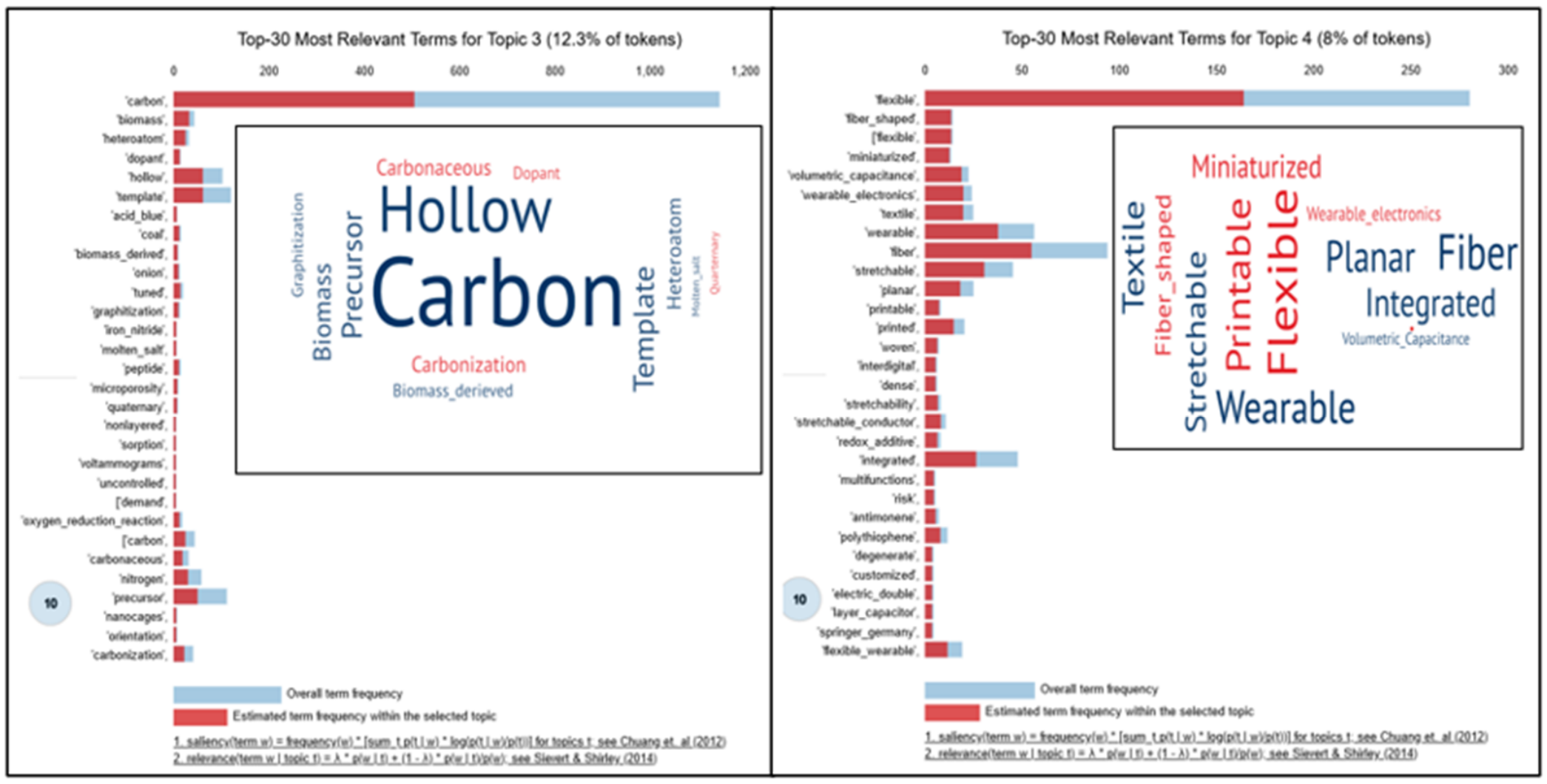

(a) (b)

Figure 5: Top words detected in (a) topic #3, (b) topic #4

### (3) Carbon Electrodes From Biomass Precursors

Words such as "Carbon," "Biomass," "Heteroatom," "Dopant," "Hollow," "Template," "Biomass-derived," "Graphitization," "Molten salt," "Quaternary," "Carbonaceous," "Nitrogen," "Precursor," and "Carbonization" constitute topic #3. Observing Fig. 5(a), we understand that this topic furnishes information about the synthesis of carbon materials for supercapacitor electrodes from biomass precursors. In addition to contributing to better cyclic stability and rate capability, the high specific surface area, hierarchical porous frameworks, and good structural stability of carbon-based materials make them the most researched material for improving the properties of supercapacitor electrode materials (27). Popular sources of carbon inclusive of graphite and coke, are non-renewable and aren't environment friendly, as a result of which the research community is seriously considering the possibility of using biomass as a precursor to carbon-based electrode materials. The biomass material is subjected to carbonization and activation to obtain the functional electrode. Carbonization involves the usage of protective gases such as Nitrogen. The activation of the biomass involves alkaline or acidic, or metal salts (28), (29) to initiate the conversion of the biochar to full-fledged electrode materials. These activating reagents induce the formation of pores of varied dimensions (macropores, mesopores, and micropores), which plays an influential role in dictating the energy storage ability of the supercapacitors (27).

Further, doping biomass with heteroatoms such as Nitrogen, Oxygen, and Sulphur increases surface properties such as wettability and electrical resistance. Precisely doping Nitrogen improves the capacitance by increasing the conductivity of carbon materials. The term "Quaternary" in the list refers to the quaternary type nitrogen in the carbon material, which is shown to improve the charge transfer through the carbon electrode (27). The term "Graphitization" in the list relates to the 'Graphitization degree' which affects the electrochemical performance of the carbon materials derived from biomass (30).

**(4) Flexible, Wearable, Printable Supercapacitors**

Demonstrated in Fig. 5(b), words such as "Flexible," "Miniatured," "Volumetric capacitance," "Wearable Electronics," "Textile," "Stretchable," "Planar," "Printable," "Woven fiber," "Flexible-wearable," "Integrated," "Interdigital" (interdigitated) hint that topic#4 deals with state of the art flexible, wearable, printable supercapacitors often integrated with textile fibers. The presence of the word 'miniatured' in this topic indicates the small-sized miniature supercapacitors, whereas flexible, stretchable, and wearable printable suggest the characteristics of the next generation supercapacitors that can be woven into textile fibers for future portable and wearable electronics (31). Words such as redox additive and Polythiophene are related to Pseudocapacitance, which is one of the two mechanisms by which charge is stored in a supercapacitor [32]. The term 'multifunctions' refers to the versatile usage of these flexible energy storage devices in medical, military, and civilian applications [32], [33]. Words such as planar and interdigital refer to the electrode architecture of the flexible supercapacitors [34]. The appearance of the word 'printable' in the research is attributed to the exploitation of additive manufacturing techniques for the fabrication of flexible supercapacitors [35].

**(5) Graphene as electrode**

Interesting fact to note in topic#5, shown in Fig. 6(a), is that the main word is "Graphene," with more than 1100 citings in this cluster, whereas the word with the next highest citing is observed for "Gel," with less than 50 citings. A few other terms, such as "Bioinspired," "Methane," and "Nanoribbon(s)," which show a handful number of citings, are also associated with Graphene. This is because while Graphene is generally synthesized using methane in the chemical vapor deposition method[36], research is also in place for its bioinspired synthesis using naturally occurring biological materials such as leaves, flower petals of plants, etc [37,38]. The word "nanoribbon" indicates that Graphene is often produced in the form of nanoscale ribbons[39]. It is not surprising since the use of Graphene as an electrode material for supercapacitors offers an array of meritorious properties such as high electrical conductivity, large specific surface area, excellent flexibility, and mechanical strength[40].

**(6) Synthesis specifications for Nanomaterials for Supercapacitor Electrodes**

The critical words of this topic include "Ice," "Stacked," "Nucleation," "Noble metals," and "Charge-carrier", as shown in Fig. 6(b). The term "Ice" indicates the usage of ice as a coolant during the synthesis of common electrode materials like rGO often fabricated as a stack of few layers or multilayers[41, 42, 43]. Often, researchers grow (nucleate) nanoparticles of noble metals like gold, silver etc. on rGO in order to improve the electrical conductivity of the rGO based electrode since noble metals are known to possess high charge-carrier density [44, 45].

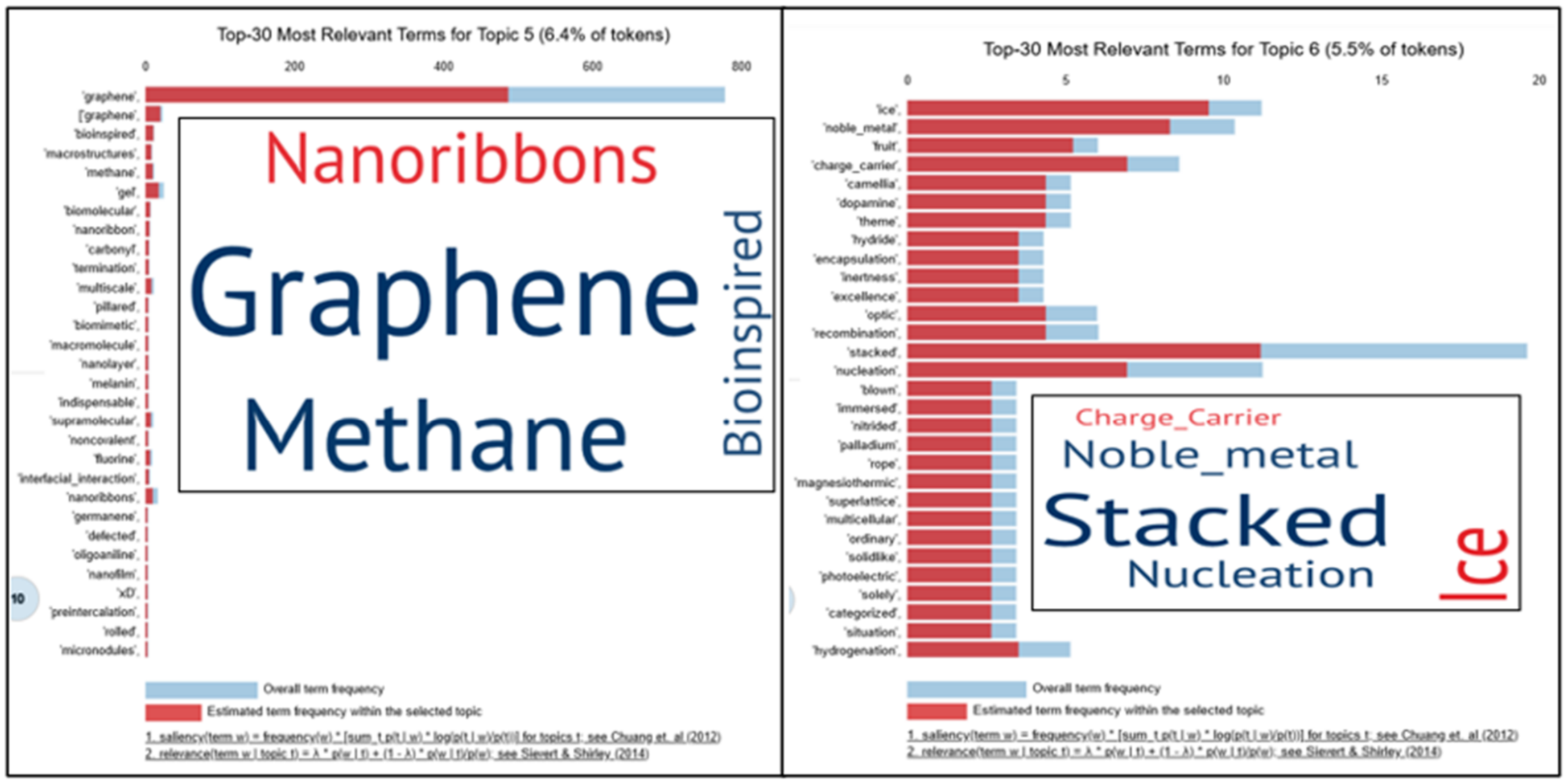

(a) (b)

Figure 6: Top words detected in (a) topic #5, (b) topic #6

**(7) Catalysis, sensing, and other applications of supercapacitor electrode materials**

As shown in Fig. 7(a), the essential words in this cluster are "Enriched," "Microtubes," "Conjoined," "Heterostructures," "Horseradish," "Peroxidase," "rGO," "Porphyrin," "Cocatalytic," "Catalyzed," "Anodization," "Banana," "Anchovy," and "Oxygen Vacancy." These words indicate that the topic revolves around the usage of supercapacitor electrode materials for sensing and catalytic applications. Macrostructures of 2D materials fabricated using tube-like morphology offers improved performance and could pave the way for filters and battery application [46]. "Conjoined" refers to the conjoined architecture of materials such as Carbon hybrid aerogels commonly used for fuel cells, supercapacitors, and sensor applications[47]. "Heterostructures" could refer to the highly prevalent usage of carbon-based heterostructures for dye-sensitized solar cells, lithium-ion batteries, supercapacitors, biosensors, etc. It is brought to the cognizance of readers that "Horseradish Peroxidase" is a single term even though the Gensim library's implementation of bigram is not able to detect the phrase. These are some of the potential pitfalls associated with our NLP model. Horseradish Peroxidase is an enzyme that finds application in catalyzing the synthesis of carbon nanomaterials-conducting polymer composites, which are majorly used in sensor and supercapacitor applications [48].

"Banana" and "Anchovies" refer to the widespread application of agro-waste-derived carbonaceous materials in sensing, dye-sensitized solar cells, electrode materials for supercapacitors, filtration, etc. [49]. "Porphyrin" refers to the extensive usage of the material in chemical sensors [50]. The word "Oxygen Vacancy" in the list hints at the influence on the electronic properties when oxygen vacancies are defect engineered into metal oxides [51].

**(8) Shapes or morphology of the electrode materials**

“Nanoplatelets”, “nanoneedle”, “rod”, “nanopowder”, “rectangular”, “microcubes”, etc., represent the shape of the particles that are often used for decorating the carbon or rGO based electrodes [7, 52]. Words such as “X-ray diffraction (XRD)”, “electron microscope”, and “spectrometry” can also be seen in Fig. 7(b), which originate from their use to characterize the shape of these nanoparticles as these shapes are identified using electron microscopy whereas their structure is identified by XRD. The morphologies of nanomaterials have a direct impact on their surface activity as the inherent porous structure of nanomaterials results in high specific surface area, a prerequisite for high specific capacitance of supercapacitors. Nanostructured design also facilitates shorter path lengths resulting in faster kinetics due to improved diffusivity of electrons and ions which in turn enhances the electrochemical activity of supercapacitor electrode systems [53, 54, 55, 56].

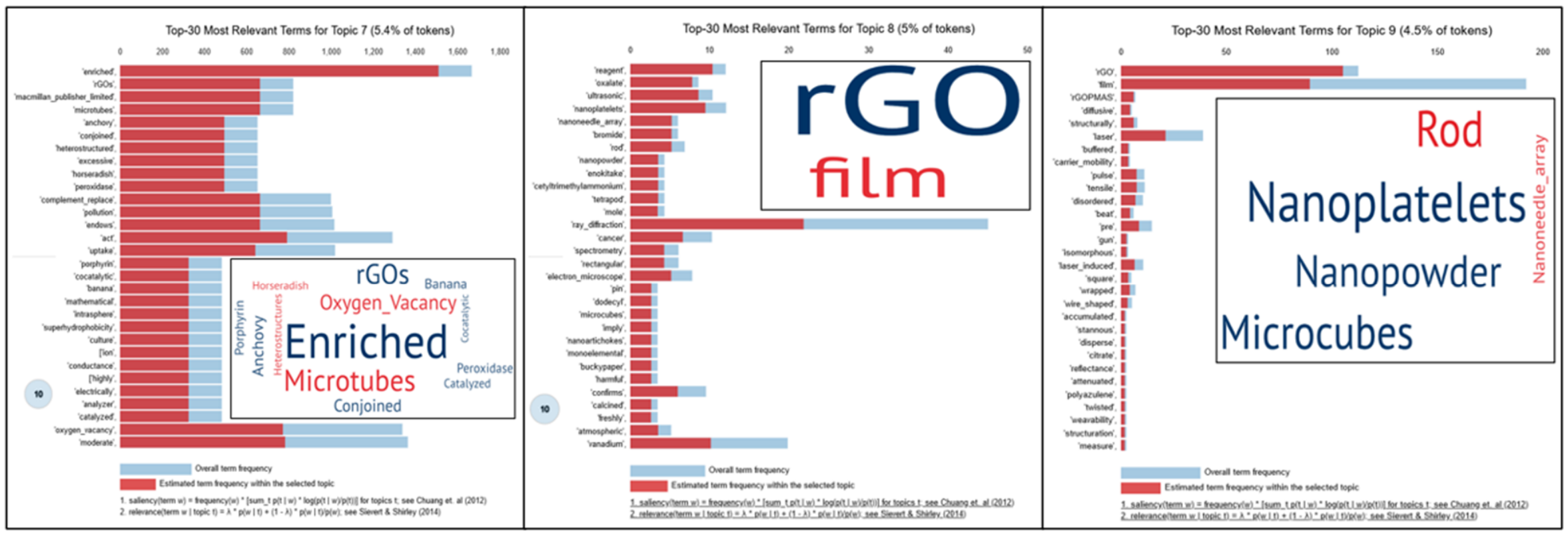

(a) (b) (c)

Figure 7: Top words detected in (a) topic #7, (b) topic #8, (b) topic #9

### (9) Reduced Graphene Oxide (rGO) as supercapacitor electrode material

As Fig. 7(c) shows, in this cluster, the main word is "Reduced Graphene Oxide," with about 1000 occurences, whereas the word with the next highest citing is observed for "Film," with about 800 citings. In the literature, often RGO is produced as a film and thus the phrase 'rGO film' is frequently used which possibly explains the nearly same number of citings of both these words. It is to be further noted that the words graphene and rGO are often used interchangeably in the literature despite the latter is the one which most people actually use for supercapacitor applications. The reason for the popularity of rGO is its ease of large scale synthesis compared to graphene without compromising much of graphene's exciting properties. Thus, in a way, both topic 5 and this topic are the similar, and may be merged together which further establishes the importance of graphene/rGO as the most important material for supercapacitor electrode. Terms such as "laser" have about one-fifth the number of citing of rGO, which assists us in safely concluding that this topic talks about the usage of rGO in fabricating supercapacitor electrodes.

### (10) Author affiliations

The words listed out by the model under the final topic are primarily associated with the authors' affiliations and id discarded as it does not yield anything concrete about supercapacitors.

## 4. Conclusion and Future Directions

The field of supercapacitors has experienced a revolution with the incorporation of nanotechnology. New nanomaterials of varying morphologies and conformations are regularly revealing beautiful chemistry and structure-function relationships. Scientific community depends on information harvested from the scientific literature for thorough and enhanced nanomaterial development for supercapacitors. Instead of this time-consuming manual process, in this work, we design an automated tool for extraction, organization and visualization of the targeted research information.

Using a natural language processing pipeline, our study provides a bird's eye view of current research trends in using nanomaterials in the emerging supercapacitor research area. An LDA-based topic modeling algorithm was implemented on the corpus to discover the nine latent topics covering characteristic and method-driven areas pertaining to the supercapacitor material research. This study is believed to contribute positively to the upcoming supercapacitor material research by highlighting the hotspot research topics which will steer the research fraternity to probe deeper and solve relevant problems. The extent and depth of our result can be compared with simple statistical analyses, like Term Frequency histogram and word-cloud that we also performed as a preliminary analysis. The comparison shows drastic improvement in terms of structured and categorised conclusions in case of the proposed methodology. The study also reduces manual labour by automatically clustering the required data in distinct topics and exhibiting a clear illustration in terms of topic distance map.

This methodology is very general and can be adopted to identify research directions for not on this, but other important material science domains as well. This intelligent and automated way of data visualization and analytics using unsupervised learning algorithms are very much appropriate for application to custom structured databases. In a future work, we would be interested to work on the figures reported in the research articles related to supercapacitors, to dig deeper with advanced image processing and deep learning tools, which is beyond the scope of this paper. We believe further works in this area will accelerate discovery and development of nanomaterials for supercapacitors.